\documentclass{PoS}

\title{Charmed baryon resonances with heavy-quark spin symmetry
\footnote{ This research was supported by
DGI and FEDER funds, under contracts FIS2011-28853-C02-02, FIS2011-24149,
FPA2010-16963 and the Spanish Consolider-Ingenio 2010 Programme CPAN
(CSD2007-00042), by Junta de Andaluc{\'\i}a grant FQM-225, by Generalitat
Valenciana under contract PROMETEO/2009/0090 and by the EU HadronPhysics2
project, grant agreement n. 227431. O.R. and L.T. wishes to acknowledge support from
the Rosalind Franklin Fellowship. L.T. acknowledges support from Ramon y Cajal
Research Programme, and from FP7-PEOPLE-2011-CIG under contract
PCIG09-GA-2011-291679.}}

\ShortTitle{Charmed baryon resonances with HQSS}

\newcommand{\SU}{{\rm SU}}
\newcommand{\U}{{\rm U}}

\newcommand{\MeV}{{\rm MeV}}

\newcommand{\ignore}[1]{} 

\author{\speaker{O.~Romanets}$^a$, L.~Tolos$^{b,a}$, C.~Garc\'ia-Recio$^c$, J.~Nieves$^d$,
  L.~L.~Salcedo$^c$, R.~G.~E.~Timmermans$^a$\\
\llap{$^a$}  KVI, University of Groningen \\
      Zernikelaan~25,~9747AA~Groningen, The~Netherlands \\
\llap{$^b$} Institut~de~Ci\`encies~de~l'Espai~(IEEC/CSIC)\\
   Campus~Universitat~Aut\`onoma~de~Barcelona, Facultat~de~Ci\`encies, Torre~C5,~E-08193~Bellaterra,~Spain \\
\llap{$^c$} Departamento~de~F\'isica~At\'omica, Molecular~y~Nuclear, and Instituto Carlos I de F{\'\i}sica Te\'orica y Computacional, Universidad~de~Granada \\
    E-18071~Granada, Spain\\
\llap{$^d$} Instituto~de~F\'isica~Corpuscular~(centro~mixto~CSIC-UV), Institutos~de~Investigaci\'on~de~Paterna\\
   Aptdo.~22085,~46071,~Valencia,  Spain \\
  E-mail: \email{o.romanets@rug.nl} }

\abstract{
Charmed baryon resonances that are generated dynamically
are studied within
 a unitary baryon-meson coupled-channel model, which incorporates
heavy-quark spin symmetry. The extension of the SU(3)
Weinberg-Tomozawa chiral Lagrangian to SU(8) spin-flavor symmetry 
plus a suitable symmetry breaking is used.  The model produces resonances with
negative parity from $s$-wave interaction of pseudoscalar and vector mesons
with $1/2^+$ and $3/2^+$ baryons. 
Some of our dynamically generated states can be readily assigned to
resonances found experimentally, while others do not have a straightforward
identification and require the compilation of more data and also refinements
of the model.
}

\FullConference{Sixth International Conference on Quarks and Nuclear Physics\\
                 April 16-20, 2012\\
                 Ecole Polytechnique, Palaiseau,  Paris}

\begin{document}

\section{Introduction}
 The observation
of new charmed and strange baryon resonances and the plausible explanation 
of their nature is an active
topic of research.
Data about such states comes from 
CLEO, Belle, BaBar and
other experiments, such as
 the planned PANDA
and CBM at the FAIR facility at GSI. Those future experiments, which involve studies of
charm physics, open the possibility for more data in the near future.  The ultimate goal is 
to understand whether those new states can
be described with the usual three-quark baryon or quark-antiquark meson
interpretation or, alternatively, qualify better as hadron molecules.

Recent approaches based on coupled-channel dynamics have proven to be quite
successful in describing the existing experimental data. In particular,
unitarized coupled-channel methods have been applied in the baryon-meson
sector with the charm degree of freedom
\cite{Tolos:2004yg,Hofmann:2005sw,Hofmann:2006qx,%
 JimenezTejero:2009vq}, partially motivated by the
analogy between the $\Lambda(1405)$ and the $\Lambda_c(2595)$. 
Other existing coupled-channel approaches are based on the
J\"ulich meson-exchange model
\cite{Haidenbauer:2007jq} or on the
hidden gauge formalism \cite{Wu:2010jy}.

However, those previous models are not consistent with heavy-quark spin
symmetry~\cite{Isgur:1989vq}, which is a proper
QCD symmetry that appears when the quark masses, such as the charm mass,
become larger than the typical confinement scale. Aiming to incorporate
heavy-quark symmetry, an SU(8) spin-flavor symmetric model has recently been
developed \cite{GarciaRecio:2008dp,Gamermann:2010zz}, which includes vector
mesons similarly to the SU(6) approach developed in the light sector of 
Ref.~\cite{GarciaRecio:2005hy}.

The objective of this work is to study dynamically
generated baryon resonances using heavy-quark spin symmetry constraints. We
focus on charm $C=1$ and strangeness $S=-2,-1$ and $0$, as well as on
sectors with $C=2$ and $3$. We therefore use the model of
Ref.~\cite{GarciaRecio:2008dp} and, as novelty, we pay special attention
to the pattern of spin-flavor symmetry breaking.
Flavor $\SU(4)$ is not a good symmetry in the limit of a heavy charm quark.
Therefore, instead of the breaking pattern $\SU(8) \supset \SU(4)$, in
this work we consider the pattern $\SU(8) \supset \SU(6)$, since the light
spin-flavor group $\SU(6)$ is decoupled from heavy-quark transformations.  This
allows us to implement heavy-quark spin symmetry in the analysis and to
unambiguously identify the corresponding multiplets among the resonances
generated dynamically~\cite{REF}.
At the same time, we are also able to assign approximate heavy
\SU(8) and light \SU(6) spin-flavor multiplet labels to the states.

\section{Framework}
For the baryon-meson interaction we use the model of
Refs.~\cite{GarciaRecio:2008dp,Gamermann:2010zz}. This model
 is based on an extension of the Weinberg-Tomozawa (WT) SU(3) chiral
Lagrangian.
The channel space includes charmed 
vector mesons and ${3/2}^+$ baryons,
in addition to pseudoscalar mesons and ${1/2}^+$ baryons.
The model obeys
spin-flavor symmetry and also heavy-quark spin symmetry (HQSS) in the
sectors studied in this work.  Schematically,
\begin{equation}
\label{symbolic}
{\mathcal L}^{SU (8)}_{ \rm WT} 
=
 \frac{1}{f^2} [[M^{\dagger} \otimes M]_{\bf 63_{a}}
  \otimes [B^{\dagger} \otimes B]_{ \bf 63} ]_{ \bf 1} 
,
\end{equation}
where $M$ is the $\bf 63$ meson multiplet and $B$ is the $\bf 120$ baryon
multiplet of \SU(8).

In the $s$-channel, the baryon-meson space reduces into four SU(8)
irreps, from which
two multiplets ${\bf 120}$ and ${\bf 168}$ are the most attractive ones, while
the ${\bf 4752}$-plet is weakly attractive and the ${\bf 2520}$-plet is
repulsive.  As a consequence, dynamically-generated baryon resonances are most
likely to occur within the ${\bf 120}$ and ${\bf 168}$ sectors. Therefore, only 
states that belong to these two representations are studied. 

To take into account the breaking of flavor symmetry introduced by the heavy
charmed quark, we consider the reduction
$\SU(8) \supset \SU(6)\times \SU_C(2) \times \U_C(1)$,
where $\SU(6)$ is the spin-flavor group for three flavors and $\SU_C(2)$ is the
rotation group of quarks with charm. We consider only $s$-wave interactions so
$ \boldmath J_C$ is just the spin carried by the charmed quarks or antiquarks.  Finally,
U$_C(1)$ is the group generated by the charm quantum number $C$.
Further, the SU(6) multiplets can be reduced under $\SU(3)\times \SU_l(2)$, the factor
$\SU_l(2)$ referring to the spin of the light quarks.
 In order to connect with the labeling $(C,S,I,J)$ based on isospin
multiplets ($S$ is the strangeness, $I$ the isospin, $J$ the spin), we further
reduce $\SU_l(2) \times \SU_C(2)\supset\SU(2)$ where $\SU(2)$ refers to the
total spin $J$, that is, we couple the spins of light and charmed quarks to
form $\SU(3)$ multiplets with well-defined $J$.

The tree-level baryon-meson interaction of the SU(8)-extended WT interaction
reads
\begin{equation}
V_{ij}(s)= D_{ij}
\frac{2\sqrt{s}-M_i-M_j}{4\,f_i f_j} \sqrt{\frac{E_i+M_i}{2M_i}}
\sqrt{\frac{E_j+M_j}{2M_j}} 
\,.
\label{eq:vsu8break}
\end{equation}
Here, $i$ and $j$ are the outgoing and incoming baryon-meson channels,
respectively.
$M_i$, $E_i$ and $f_i$ stand, respectively, for the mass, 
the center-of-mass energy of the baryon, 
and the meson decay constant in the $i$ channel.
 $D_{ij}$ are the matrix elements coming from the SU(8) group structure of the 
coupling for the various $CSIJ$ sectors. 
In order to calculate the scattering amplitudes, $T_{ij}$, we solve the
on-shell Bethe-Salpeter equation in coupled channels using the interaction
matrix $V$ as kernel:
\begin{equation}
\label{LS}
 T(s) =\frac{1}{1-V(s)G(s)} V(s).
\end{equation}
$G(s)$ is a diagonal matrix containing the baryon-meson propagator for each
channel. The quantities $D$, $T$, $V$, and $G$ are matrices in coupled-channel space.
The loop function $G^{0}_{ii}(s)$ is logarithmically ultraviolet
divergent and is regularized by means of the subtraction point 
regularization \cite{Nieves:2001wt}.

The dynamically-generated baryon resonances can be obtained as poles of the
scattering amplitudes.
The mass and the width of the resonance can be found from the position of the
pole on the complex energy plane. Close to the pole, the $T$-matrix behaves as
\begin{equation}
\label{Tfit}
 T_{ij} (s) \approx \frac{g_i g_j}{\sqrt{s}-\sqrt{s_R}}
\,,
\end{equation}
and in this way the coupling constants $g_i$ to different meson-baryon channels
can be found. 

The matrix elements $D_{ij}$ display exact SU(8) invariance, but this symmetry
is severely broken in nature, so we implement a symmetry-breaking
mechanism. The symmetry breaking pattern, with regards to flavor, follows the chain
$\SU(8)\supset\SU(6)\supset\SU(3)\supset\SU(2)$, where the last group refers
to isospin. 
The symmetry breaking is introduced
by means of a deformation of the mass and decay constant
parameters. 
This allows us to
assign well-defined SU(8), SU(6), and SU(3) labels to the resonances and to
find HQSS invariant states.

\section{ Dynamically generated charmed and strange baryon states }
Our model obtains the dynamically generated states in different charm and
strange sectors, namely $C=1,2,3$ and the corresponding 
strangeness numbers~\cite{REF}. We have assigned to some of them a
tentative identification with known states from the PDG
\cite{Nakamura:2010zzi}. This identification is made by comparing the data
from the PDG on these states with the information we extract from the poles,
namely the mass, width and, most important, the couplings. The couplings give
us valuable information on the structure of the state and on the possible
decay channels and their relative strength.

One of the sectors that we study is $C=1,~S=0,~I=0$, which corresponds to 
$\Lambda_c$ spin-$1/2$ and spin-$3/2$ states. Results for
$C=1$ and $S=0$ were reported previously in
Ref.~\cite{GarciaRecio:2008dp}. However, the analysis of the dynamically
generated states in terms of the attractive $\SU(8) \supset \SU(6) \supset
\SU(3) \supset \SU(2)$ multiplets was not done in this previous reference.
Here we are able to assign SU(8), SU(6), and SU(3) labels to the resonances.
Simultaneously, we also classify the resonances into HQSS multiplets, in
practice doublets and singlets. 
 We obtain
the three lowest-lying states of Ref.~\cite{GarciaRecio:2008dp} in this
sector. However, those states appear with slightly different masses due 
to the different subtraction point, and different $D_s$ and $D_s^*$ meson decay 
constants. The experimental $\Lambda_c(2595)$ resonance can be identified with the 
${\bf  21}_{2,1}$ pole 
\footnote{ $R_{2J_c+1,C}$ stands for $R$ irrep
 of SU(6).}
that we found around $2618.8\,\MeV$, as similarly done in
Ref.~\cite{GarciaRecio:2008dp}. The width in our case is, however,
smaller than the experimental value, but
we have not included the three-body decay
channel $\Lambda_c \pi \pi$, which already represents almost one third of the
decay events \cite{Nakamura:2010zzi}.
We also observe a second broad $\Lambda_c$ resonance at 2617~MeV with 
a large coupling to the open channel $\Sigma_c \pi$, very
close to $\Lambda_c(2595)$. This is precisely the same two-pole pattern found
in the charmless $I = 0, S = -1$ sector for the $\Lambda(1405)$
\cite{Jido:2003cb}. The third spin-$1/2$
$\Lambda_c$ resonance with a mass of 2828~MeV  is mainly originated by 
a strong attraction in the $\Xi_c K$
channel but it cannot be assigned to any experimentally known resonance.
We also find one spin-$3/2$ resonance in this sector located at $(2666.6 -i 26.7\,\MeV)$.
Similarly to the Ref.~\cite{GarciaRecio:2008dp} this resonance is assigned 
to the experimental $\Lambda_c(2625)$. A similar resonance was found at $2660\,\MeV$ 
in the $t$-channel vector-exchange model of Ref.~\cite{Hofmann:2006qx}. 
The novelty of our calculation
is that we obtain a non-negligible contribution from the baryon-vector meson
channels to the generation of this resonance, as already observed in
Ref.~\cite{GarciaRecio:2008dp}.
 
The three $\Sigma_c$ resonances obtained for $C=1, S=0, I=1, J=1/2$ 
with masses 2571.5, 2622.7 and 2643.4~MeV and widths 0.8, 188.0 and 
87.0~MeV respectively 
are predictions of our model, since no experimental data
have been observed in this energy region. Our predictions here nicely agree
with the three lowest lying resonances found in
Ref.~\cite{GarciaRecio:2008dp}. 
The model of Ref.~\cite{JimenezTejero:2009vq}
predicts the existence of two resonances with
$C=1,~S=0,~I=1,~J=\frac{1}{2}$. In this reference, the first one has a mass of
$2551\,\MeV$ with a width $0.15\,\MeV$. It couples strongly to the $\Sigma
D_s$ and $N D$ channels and, therefore, might be associated with the resonance
$\Sigma_c(2572)$ with $\Gamma=0.8\,\MeV$ of our model. Nevertheless, in our
model this resonance couples most strongly to the other channels which
incorporate vector mesons, such as $\Sigma^* D_s^*$ and $\Delta D^*$.
The second resonance predicted in Ref.~\cite{JimenezTejero:2009vq} has a mass
of $2804\,\MeV$ and a width of $5\,\MeV$, and it cannot be compared to any of
our results.
We predict two spin-$3/2$ $\Sigma_c$ resonances.
The first one, a bound state at 2568.4~MeV, 
lies below the threshold of any possible 
decay channel. This state is thought to be the charmed counterpart of the
hyperonic $\Sigma(1670)$ resonance. While the $\Sigma(1670)$ strongly couples
to $\Delta \bar{K}$ channel, this resonance is mainly generated by the
analogous $\Delta D$ and $\Delta D^*$ channels.
The second state at $2692.9 -i 33.5$ ~MeV has no direct comparison with the
available experimental data.

We also study the $C=1$, $S=-1$, $I=1/2$ sector for spin $J=1/2$ and
$J=3/2$. Those states are
labeled by $\Xi_c$ and our model predicts the existence of nine states
stemming from the strongly attractive {\bf 120} and {\bf 168} SU(8)
irreducible representations. 
Among them six states have spin $J=1/2$.
 In the energy range where these six states predicted by
our model lie, three experimental resonances have been seen by the Belle, E687, and
CLEO Collaborations: $\Xi_c(2645)~ J^P=3/2^+$ \cite{Nakamura:2010zzi,
  Lesiak:2008wz},
$\Xi_c(2790)~J^P=1/2^-$ \cite{Nakamura:2010zzi, Csorna:2000hw} and
$\Xi_c(2815)~J^P=3/2^-$
\cite{Nakamura:2010zzi,Lesiak:2008wz}. While $\Xi_c(2645)$
cannot be identified with any of our states due to the
parity, the $\Xi_c(2790)$ might be assigned with one of the six resonances in
the $J=1/2$ sector.
The state $\Xi_c(2790)$ has a width of $\Gamma<12-15\,\MeV$ and it decays to
$\Xi_c' \pi$, with $\Xi_c' \rightarrow \Xi_c \gamma$. 
We assign it to the
$2804.8 -i 13.5$~MeV state found in our model because of the large 
$\Xi'_c\pi$ coupling and the
fact that a slight modification of the subtraction point can lower its
position to $2790\,\MeV$ and most probably reduce its width as it will get
closer to the $\Xi_c' \pi$ channel, the only channel open at those energies
that couples to this resonance. Moreover, this seems to be a reasonable
assumption in view of the fact that, in this manner, this $\Xi_c$ state is the
HQSS partner of the $\Xi_c^*(2845)$ state, which we will identify with the
$\Xi_c^*(2815)$ resonance of the PDG. It could be also possible to
identify our pole at $2733\,\MeV$ from the {\bf 168} irreducible
representation with the experimental $\Xi_c(2790)$ state. In that case, one
might expect that if the resonance position gets closer to the physical mass
of $2790\,\MeV$, its width will increase and it will easily reach values of
the order of $10\,\MeV$.
As it was mentioned above, 
the only experimental $J^P=3/2^-$ baryon resonance with a mass in the energy
region of interest is $\Xi_c(2815)$. The full width is
expected to be less than $3.5\,\MeV$ for $\Xi_c^+(2815)$ and less than
$6.5\,\MeV$ for $\Xi_c^0(2815)$, and the decay modes are $\Xi_{c+} \pi^+
\pi^-$, $\Xi_{c0} \pi^+ \pi^-$. We obtain two resonances at $2819.7 -i 16.2\,\MeV$ and
$2845.2 -i 22.0\,\MeV$, respectively, that couple strongly to $\Xi_c^* \pi$, with
$\Xi_c^* \rightarrow \Xi_c \pi$. Allowing for this possible indirect
three-body decay channel, we might identify one of them to the experimental
result. This assignment is, indeed, possible for the state at $2845.2\,\MeV$
if we slightly change the subtraction point. In this way, we will lower its
position and reduce its width as it gets closer to the open $\Xi_c^* \pi$
channel.
 
According to our analysis, in the $C=1, S=-2, I=0, J=1/2$ sector
there are three bound states ($\Omega_c$) with masses $2810.9$, $2884.$5 and 
$2941.6$~MeV.
There is no experimental information on those excited states. However, our
predictions are comparable to recent calculations of
Refs.~\cite{JimenezTejero:2009vq,Hofmann:2005sw}.
In both these references, vector baryon-meson channels were not considered,
breaking in this manner HQSS. In fact,  it is worth noticing that the coupling
to vector baryon-meson states plays an important role in the generation of the
baryon resonances in this sector.
Further, we obtain two spin-$3/2$ bound  states $\Omega_c$
with masses $2814.3$ and
$2980.0$~MeV, which mainly couple to $\Xi D^*$ and $\Xi^* D^*$, and to $\Xi_c^* \bar K$,
respectively. As in the $J=1/2$ sector, no experimental information is
available.

We also obtain baryon resonances with $C=2$ and $3$ (see Ref.~\cite{REF}). The appropriate
numbers of strangeness of the states with charm 2 are 0~($\Xi_{cc}$) and 
$-1$~($\Omega_{cc}$), which result from the group 
decomposition of the $\bf 120$ and $\bf 168$ SU(8) representations.
Finally, the $\Omega_{ccc}$ states with $C=3, S=0, I=0$ are studied. 
 At the moment no experimental information is 
available for those ones. To our knowledge, these are the first predictions
in these sectors deduced from a model fulfilling HQSS. 

\section{Summary}
In the present work, we have studied odd-parity charmed baryon resonances
within a coupled-channel unitary approach that implements the characteristic
features of HQSS.
 This is
accomplished by extending the $\SU(3)$ WT chiral interaction to $\SU(8)$
spin-flavor symmetry and implementing a strong flavor symmetry breaking.
 We have
discussed the predictions of the model for all $C=1$ strange sectors and have
also looked at the $C=2$ and $3$ predicted states.
 We have restricted our study 
to the 288 states (counting multiplicities in spin and isospin) that stem from
the attractive {\bf 168} and {\bf 120} representations, for which we believe the
predictions of the model are more robust.
To identify these states, we have adiabatically followed the trajectories of
the {\bf 168} and {\bf 120} poles, generated in a symmetric $\SU(8)$ world,
when the symmetry is broken down to $\SU(6) \times \SU_C(2)$ and later
$\SU(6)$ is broken down to $\SU(3) \times \SU(2)$.
A first result of this work is that we have been able to identify
the {\bf 168} and {\bf 120} resonances among the plethora of resonances
predicted in Ref.~\cite{GarciaRecio:2008dp} for the different strangeness
$C=1$ sectors.
 Thus, we interpret the $\Lambda_c(2595)$ and
$\Lambda_c(2625)$ as a members of the $\SU(8)$ {\bf 168}-plet, and in both
cases with a dynamics strongly influenced by the $ND^*$ channel, in sharp
contrast with previous studies inconsistent with HQSS.
 There is scarce experimental information in all studied
sectors. 
Other identifications correspond to the three-star $\Xi_c(2790)$ and
$\Xi_c(2815)$ resonances in the $C=1, S=-1, I=1/2$ sectors. 
We believe that the rest of our predictions are
robust, and will find experimental confirmation in the future. 
In particular, the program of PANDA of FAIR are
of particular relevance.



\begin{thebibliography}{99}

\bibitem{Tolos:2004yg}
  L.~Tolos, J.~Schaffner-Bielich and A.~Mishra,
  Phys.\ Rev.\  C {\bf 70}, 025203 (2004);
  M.~F.~M.~Lutz and E.~E.~Kolomeitsev,
  Nucl.\ Phys.\  A {\bf 730}, 110 (2004);
  T.~Mizutani and A.~Ramos,
  Phys.\ Rev.\ C {\bf 74}, 065201 (2006).




\bibitem{Hofmann:2005sw} 
  J.~Hofmann and M.~F.~M.~Lutz,
  Nucl.\ Phys.\ A {\bf 763}, 90 (2005).

\bibitem{Hofmann:2006qx}
  J.~Hofmann and M.~F.~M.~Lutz,
  Nucl.\ Phys.\  A {\bf 776}, 17 (2006).



\bibitem{JimenezTejero:2009vq} 
  C.~E.~Jimenez-Tejero, A.~Ramos and I.~Vidana,
  Phys.\ Rev.\ C {\bf 80}, 055206 (2009).

\bibitem{Haidenbauer:2007jq}
  J.~Haidenbauer, G.~Krein, U.~G.~Meissner and A.~Sibirtsev,
  Eur.\ Phys.\ J.\  A {\bf 33}, 107 (2007);
  J.~Haidenbauer, G.~Krein, U.~G.~Meissner and L.~Tolos,
  Eur. Phys. J A {\bf 47}, 18 (2011)




\bibitem{Wu:2010jy}
  J.~-J.~Wu, R.~Molina, E.~Oset and B.~S.~Zou,
  Phys.\ Rev.\ Lett.\  {\bf 105}, 232001 (2010);

\bibitem{Isgur:1989vq}
  N.~Isgur, M.~B.~Wise,
  Phys.\ Lett.\  {\bf B232}, 113 (1989);
  M.~Neubert,
  Phys.\ Rept.\  {\bf 245}, 259-396 (1994).



\bibitem{GarciaRecio:2008dp} 
  C.~Garcia-Recio  {\it et al.},
  Phys.\ Rev.\ D {\bf 79}, 054004 (2009).

\bibitem{Gamermann:2010zz} 
  D.~Gamermann  {\it et al.},
  Phys.\ Rev.\ D {\bf 81}, 094016 (2010).

\bibitem{GarciaRecio:2005hy} 
  C.~Garcia-Recio, J.~Nieves and L.~L.~Salcedo,
  Phys.\ Rev.\ D {\bf 74}, 034025 (2006);
  H.~Toki, C.~Garcia-Recio and J.~Nieves,
  Phys.\ Rev.\  D {\bf 77}, 034001 (2008).


\bibitem{REF}
  O.~Romanets {\it et al.} ( Phys.\ Rev.\ D, in press),
 [hep-ph/1202.2239].

\bibitem{Nieves:2001wt} 
  J.~Nieves and E.~Ruiz Arriola,
  Phys.\ Rev.\ D {\bf 64}, 116008 (2001).

\bibitem{Nakamura:2010zzi} 
  K.~Nakamura {\it et al.}  [Particle Data Group Collaboration],
  J.\ Phys.\ G G {\bf 37}, 075021 (2010).

\bibitem{Jido:2003cb} 
  D.~Jido, J.~A.~Oller, E.~Oset, A.~Ramos and U.~G.~Meissner,
  Nucl.\ Phys.\ A {\bf 725}, 181 (2003).

\bibitem{GarciaRecio:2002td}
  C.~Garcia-Recio, J.~Nieves, E.~Ruiz Arriola and M.~J.~Vicente Vacas,
  Phys.\ Rev.\ D {\bf 67}, 076009 (2003).

\bibitem{Lesiak:2008wz}
  T.~Lesiak {\it et al.} [ Belle Collaboration ],
  Phys.\ Lett.\  {\bf B665}, 9-15 (2008).

\bibitem{Csorna:2000hw}
  S.~E.~Csorna {\it et al.} [ CLEO Collaboration ],
  Phys.\ Rev.\ Lett.\  {\bf 86}, 4243-4246 (2001).

\end{thebibliography}
\end{document}